\documentclass[a4paper,aps,english,preprint]{revtex4-1}
 \usepackage[T1]{fontenc}
\usepackage{color}
\usepackage{float}
\usepackage{amsmath}
\usepackage{graphicx}
\usepackage{esint}

\makeatletter

\pdfpageheight\paperheight
\pdfpagewidth\paperwidth

\newcommand{\lyxdot}{.}

\@ifundefined{textcolor}{}
{%
 \definecolor{BLACK}{gray}{0}
 \definecolor{WHITE}{gray}{1}
 \definecolor{RED}{rgb}{1,0,0}
 \definecolor{GREEN}{rgb}{0,1,0}
 \definecolor{BLUE}{rgb}{0,0,1}
 \definecolor{CYAN}{cmyk}{1,0,0,0}
 \definecolor{MAGENTA}{cmyk}{0,1,0,0}
 \definecolor{YELLOW}{cmyk}{0,0,1,0}
}

\@ifundefined{date}{}{\date{}}
\makeatother

\renewcommand{\min}{{\hbox{min}}  }

 \usepackage{babel}
\begin{document}

\title{Topological Defects and Defects-free states in toroidal nematics}

\author{Han Miao} 
\affiliation{ Department of Physics, Shanghai Jiao Tong University, Shanghai
200240 China}

\author{Yao Li}
\affiliation{ Department of Physics, Tsinghua University, Beijing 100084,
China}
\affiliation{ Department of Physics, Shanghai Jiao Tong University, Shanghai
200240 China}

\author{Hongru Ma}
\email{hrma@sjtu.edu.cn}
\affiliation{The State Key Laboratory Of Metal Matrix Composites}
\affiliation{ Key Laboratory of Artificial Structures and Quantum Control (Ministry of Education)}  
\affiliation{School of Mechanical Engineering, Shanghai Jiao Tong University,
Shanghai 200240 China}

\begin{abstract}
We investigated the nematic ordering on a torus by means of analytic method and the method
of simulated annealing, the Frank free energy, both in the standard form and covariant form, were used in the study. The defect free state was found to be the ground state in both cases. However, in the case of the standard model, there are two kinds of defective free ordering and a transition between the two occurs at a critical value of radius ratio $k=\frac{r}{R}$. The first one is $\theta=0$ in the small $k$ regime and the second one is a variable  $\theta$ with position of the torus. In the case of the covariant model the ground state is confirmed to be the infinitely degenerate of $\theta$ equals to a random constant. The states with defects are the excited states, where the pairs of defects excited and, duo to the barrier between positive and negative defects, have pretty long life. The behavior of the defect state basically the same for both of the two models.  
\end{abstract}
\maketitle

\section{Introduction}

Interplays of order and geometry play fundamental roles in many physical
systems\citep{Nelson2002a,Bowick2009}. In-plane order on a two-dimensional
curved surface attracted research interests in recent years since the 
discoveries of ordered phases $L_{\beta}$ and $P_{\beta}$  in phospholipid
membranes\citep{nelson2004statistical}. After the pioneer work by Nelson and Peliti\citep{Nelson1987}
   on interplay of crystalline and hexatic order on
fluctuating membranes,  quite a lot of researches are done for understanding
the relation between in-plane order and the geometry where it live
on.

Lubensky and Prost \citep{Lubensky1992} showed that the equilibrium
positions of topological defects for smectic-C order on a spherical
vesicle are on two opposite poles of the sphere, while the topological
defects of the nematic order and hexatic order on a spherical vesicle
stay at the corner of a tetrahedron and an icosahedron, respectively,
and Park, Lubensky and MacKintosh\citep{Park1992} extended this study
to deformable spherical vesicles. Nelson\citep{Nelson2002} proposed
a possible application of these defects as attachment points for chemical
linkers, % resulting to colloids or
%droplets with determinative number and pattern of valence,
suitable for molecular self-assembly. One example of this promising
route of designing superstructure materials has been achieved for
divalent case by Smectic-C tilt order \citep{DeVries2007}. In recent
years, a number of experimental, theoretical and simulation studies
have been dedicated to the issue of nematic and hexatic order on spherical
surfaces \citep{Liang2011,Lopez-Leon2011,Shin2008,Bates2008,Zhang2012a,Zhang2012}.
%this part need to be reduced.

Besides the intensive studies of spherical topology surfaces, vesicles
of toroidal topology gave rise to interest after prediction and first
experiment observation of toroidal vesicles\citep{Ou-Yang1990,Mutz1991}
more than 20 years ago. Evans first gave a theoretical discussion
on covariant model of $p$-atic order on a toroidal membrane\citep{Evans1995},
revealing topological defects may appear in ground state even not
required by topological constraints. Then Bowick, Nelson and Travesset
investigated hexatic order in toroidal geometries more extensively\citep{Bowick2004}.
And $3D$ derivative XY Model on toroidal surface was studied by Seilinger
et. al.\citep{Selinger2011} using Monte Carlo methods very recently.

In this paper, we describe nematic order on toroidal surface by studies
of two different models, covariant derivative and $3D$ derivative
Franck free energy models,   depending on
the molecular details of a real system, both of the two models can be 
approximate models for real systems. These models can be used to describe toroidal
vesicles self-assembly by amphiphilic liquid crystalline side-chain
block copolymers, where experimental examples of the spherical topology
were achieved by Jia et. al.\citep{Jia2009,Jia2011}. By using both
numerical and Monte Carlo methods, we investigate these two models,
exploring the defect-free and defective structures for toroidal surfaces
at various parameters extensively. 
%{ \color{red} We give the comparison of results
%of covariant derivative and $3D$ derivative models, both the defect-free
%states and defects behavior, and analysis the difference and extension
%of previous relative studies.}

\section{Model and Methods\label{sec:Model-and-Methods}}
\begin{figure}[tbh]
\includegraphics[width=0.6\paperwidth]{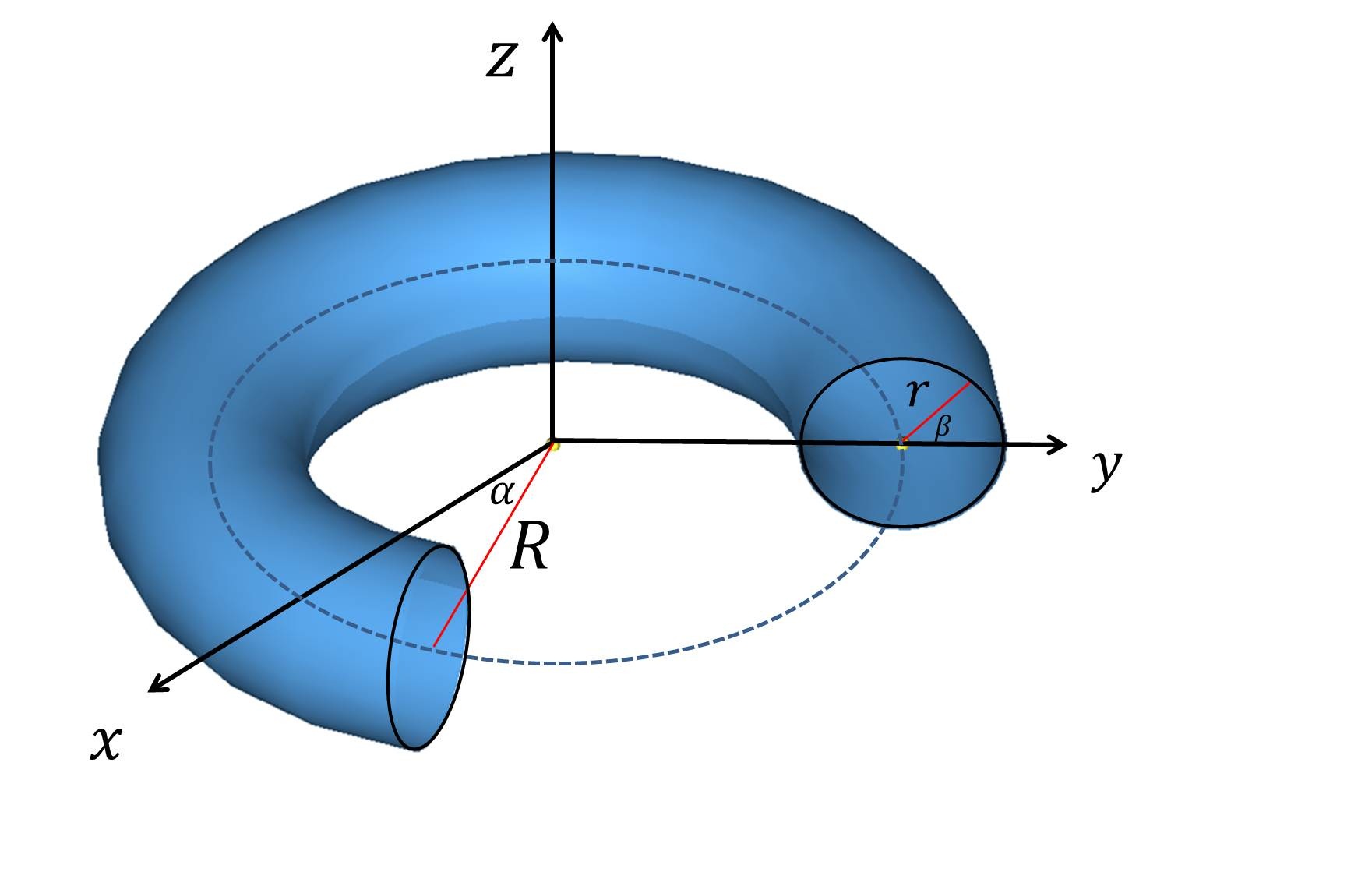}
\caption{The coordinate system and parametrization of a circular torus of radii $R$ and $r$.\label{fig:torus-Parametrization}}
\end{figure}

The  purpose of this study is to investigate the nematic orders on the torus, thus we first define our problem and give the method of study in detail. The torus to be studied
is a circular torus with radius $R$, the distance between the center of the torus and the center
of the tube, and with tube radius $r$.  In fact, here only the ratio $R/r$ is an independent variable, however, we keep the two for clarity of the following presentation. The coordinate system is chosen in the following, the $z$ axis is perpendicular to the plane of the large circle
of the torus, and $xy$ plane is the plane of the circle, the origin of the coordinate system is
taken to be the center of the torus.  In this system of coordinates, a point on the torus can be specified by two parameters $\alpha$ and $\beta$, both varies from $0$ to $2\pi$, represent
the angle of the projection to $xy$ plane of the position vector with respect to $x$ axis and the local  angle of the point  with respect to the $xy$ plane  when taking the intersection of the
projection and the center circle of the tube as the origin.  A sketch of the coordinate system is
shown in Fig.\ref{fig:torus-Parametrization}. With the two parameters $\alpha$ and $\beta$, the coordinates of a point on the torus is given by
\begin{equation}
\boldsymbol{X}=\left(\begin{array}{c}
\cos\alpha\left(R+r\cos\beta\right)\\
\sin\alpha\left(R+r\cos\beta\right)\\
r\sin\beta
\end{array}\right)
\end{equation}
  The in-plane nematic liquid crystal order is modeled by a director field $\hat n(\mathbf x)$, which is a
unit headless director field  living in the tangent plane of the toroidal
surface. Similar to  a previous study for tilt order
on curved surface by J. V. Selinger et al  Ref.\citealp{Selinger2011},  the nematic elastic energy is taken to be  the standard Frank free energy form
\begin{equation}
F=\int K\left(\nabla\hat{n}\right)^{2}dA\label{eq:3D-Continue}
\end{equation}
where $\nabla\hat{n}$ is the three-dimensional (3D) gradient of headless
director field $\hat{n}$.     For comparison,
we also consider the covariant gradient Frank free energy model\citep{Evans1995}
\begin{equation}
F=\int K\left(D\hat{n}\right)^{2}dA\label{eq:Covarint-Continue}
\end{equation}
where $D\hat{n}$ denotes the covariant gradient of $\hat{n}$.

The discretization is necessary in order to do numerical simulations, in this
calculation the toroidal surface is discretized
into a triangular lattice. For each triangle $i$, we assign an in-plane
unit vector $\hat{n}_{i}$ to represent its local in-plane liquid
crystal order. The  unit vector pairs  $\hat{n}_{i}$ and $\hat{n}_{j}$  interact with each other only when the triangles $i$ and $j$  are nearest neighbors. For the
3D gradient model,  the Frank free energy  on the discretized surface is given by the
following expression
\begin{equation}
F=2K\underset{\left\langle ij\right\rangle }{\sum}S_{ij}d_{ij}^{-2}\left\{ 1-\left[\hat{n}_{i}\cdot\hat{n}_{j}\right]^{2}\right\}.  \label{eq:3D-discrete}
\end{equation}
Here $d_{ij}$
is the bond length connecting $v_{i}$ and $v_{j}$ where $v_{i}$
denotes the center of the triangle $i$.  $S_{ij}=S_{\Delta v_{i}ab}+S_{\Delta v_{j}ab}$,
where $S_{\Delta v_{i}ab}$ and $S_{\Delta v_{j}ab}$ are  the areas  of the two triangles with vertices $(v_i, a, b)$ and $(v_j, a, b)$ respectively,
$a$ and $b$ are the two end points of the edge shared by triangles $i$ and $j$.
When the area $S_{ij}$ tends to zero the discretized form approaches to the correct   continuum limit \ref{eq:3D-Continue} as expected.

 For the covariant gradient model Eq.\ref{eq:Covarint-Continue},
the discrete free energy form is given by
\begin{equation}
F=2K\underset{\left\langle ij\right\rangle }{\sum}S_{ij}d_{ij}^{-2}\left\{ 1-\left[\hat{n}_{i}\cdot\Gamma\left(j,i\right)\hat{n}_{j}\right]^{2}\right\} \label{eq:covariant-discrete}
\end{equation}
where $\Gamma\left(j,i\right)\hat{n}_{j}$ is the  parallel
transport  operation that transports  the vector $\hat{n}_{j}$ from triangle $j$ to $i$,  given by\citep{Ramakrishnan2010}
\begin{equation}
\Gamma\left(j,i\right)\hat{n}_{j}=\left(\hat{n}_{j}\cdot\hat{\zeta}_{ji}\right)\hat{\zeta}_{ij}+\left[\hat{n}_{j}\cdot\left(\hat{e}_{j}\times\hat{\zeta}_{ji}\right)\right]\left(\hat{e}_{i}\times\hat{\zeta}_{ij}\right)
\end{equation}
where $\zeta_{ij}=\mathbf{P}_{i}\hat{r}_{ij}$ is the best estimate
for the directions of geodesic connecting the center of triangle $i$
and $j$, here $\hat{r}_{ij}$ is the direction vector connecting them
and $\mathbf{P}_{i}$ is the plane projector of triangle $i$.

Now we describe our simulation method. The ordering configuration is determined by the minimization of the free energy numerically in the discrete form, however, direct minimization is impractical because the number of variables is very large.  Thus the simulated annealing Monte Carlo method is employed for the minimization
of the free energy. The initial configuration is the randomly assigned nematic vectors  to each triangles, then   the nematic vector of
a random picked triangle is randomly rotated within the tangent plane.
The acceptance probability  for each move is given
by the Metropolis rule
\begin{equation}
P_{acc}=min\left\{ 1,exp\left(-\Delta F/k_{b}T\right)\right\}
\end{equation}
where $\Delta F$ is the free energy difference before and after a
 move. The initial temperature is high enough and then lowered in every
 5000 Monte Carlo steps. When the temperature is low enough so that a definite configuration
 emerges, the calculation stopped.

\section{Results and discussion\label{sec:Results-and-discussion}}

\subsection{Defect-free states}

According to Poincar\'{e}-Hopf theorem, the sum of the winding numbers
of all in-plane order defects equals to the Euler characteristic number
$\chi$ of the topological surface they live in. Particularly for
toroidal surface, with Euler characteristic number $\chi=0$, the
net winding number of all topological defects should be zero.

First we consider the defect-free states for 3D gradient model described by  equation \ref{eq:3D-Continue}.
The two perpendicular tangent vectors of the toroidal surface are given by
\begin{equation}
\begin{split}\boldsymbol{X}_{\alpha}=\left(\begin{array}{c}
-\sin\alpha\left(R+r\cos\beta\right)\\
\cos\alpha\left(R+r\cos\beta\right)\\
0
\end{array}\right)\\
\boldsymbol{X}_{\beta}=\left(\begin{array}{c}
-r\cos\alpha\sin\beta\\
-r\sin\alpha\sin\beta\\
r\cos\beta
\end{array}\right)
\end{split}
\end{equation}
and the covariant form of the metric tensor is evaluated from the tangent vectors as:
\begin{equation}
g_{ij}=\left(\begin{array}{cc}
\boldsymbol{X}_{\alpha}\cdot\boldsymbol{X}_{\alpha} & \boldsymbol{X}_{\alpha}\cdot\boldsymbol{X}_{\beta}\\
\boldsymbol{X}_{\alpha}\cdot\boldsymbol{X}_{\beta} & \boldsymbol{X}_{\beta}\cdot\boldsymbol{X}_{\beta}
\end{array}\right)=\left(\begin{array}{cc}
\left(R+r\cos\beta\right)^{2} & 0\\
 0 & r^{2}
\end{array}\right)
\end{equation}
with is corresponding contravariant  form
\begin{equation}
g^{ij}=\left(\begin{array}{cc}
\frac{1}{\left(R+r\cos\beta\right)^{2}} & 0\\
0 & \frac{1}{r^{2}}
\end{array}\right)\label{eq:g^ij}
\end{equation}
\begin{equation}
\sqrt{g}=\det(g_{ij})=r\left(R+r\cos\beta\right)\label{eq:det-g}
\end{equation}
The nematic  vector can be  represented by its orientation angle $\theta$ in the tangent space
by
\begin{equation}
\hat{\boldsymbol{n}}=\cos\theta\hat{\boldsymbol{X}}_{\alpha}+\sin\theta\boldsymbol{\hat{X}}_{\beta}
\end{equation}
where the orientation angle is a function of the parameters  $\alpha$ and $\beta$, $\theta= \theta\left(\alpha,\beta\right)$, and
the normalized tangent vectors are
\begin{equation}
\begin{split}\boldsymbol{\hat{X}}_{\alpha}=\left(\begin{array}{c}
-\sin\alpha\\
\cos\alpha\\
0
\end{array}\right)\\
\boldsymbol{\hat{X}}_{\beta}=\left(\begin{array}{c}
-\cos\alpha\sin\beta\\
-\sin\alpha\sin\beta\\
\cos\beta
\end{array}\right)
\end{split}
\end{equation}
The free energy (\ref{eq:3D-Continue}) can be explicitly parametrized in this representation as
\begin{equation}
F/K=\oint_{torus}\left(\nabla\hat{n}\right)^{2}dS=\int\int\sqrt{g}\underset{i,j=\alpha,\beta}{\sum}g^{ij}\partial_{i}\hat{\boldsymbol{n}}\cdot\partial_{j}\hat{\boldsymbol{n}}d\alpha d\beta\label{eq:3D-explicit}
\end{equation}
The derivatives of the nematic vectors  are
\begin{equation}
\begin{split}\partial_{\alpha}\hat{\boldsymbol{n}}=\cos\theta\boldsymbol{\hat{X}}_{\alpha\alpha}-\sin\theta\theta_{\alpha}\hat{\boldsymbol{X}}_{\alpha}+\sin\theta\boldsymbol{\hat{X}}_{\beta\alpha}+\cos\theta\theta_{\alpha}\boldsymbol{\hat{X}}_{\beta}\\
\partial_{\beta}\hat{\boldsymbol{n}}=\cos\theta\boldsymbol{\hat{X}}_{\alpha\beta}-\sin\theta\theta_{\beta}\hat{\boldsymbol{X}}_{\alpha}+\sin\theta\boldsymbol{\hat{X}}_{\beta\beta}+\cos\theta\theta_{\beta}\boldsymbol{\hat{X}}_{\beta}
\end{split}
\label{eq:d-nematic}
\end{equation}
In the above expression the subscript means the partial differentiation with respect to the variable. The derivatives of the unit  tangent vectors are given by
\begin{equation}
\begin{split}\boldsymbol{\hat{X}}_{\alpha\alpha}= & \left(\begin{array}{c}
-\cos\alpha\\
-\sin\alpha\\
0
\end{array}\right)\\
\boldsymbol{\hat{X}}_{\alpha\beta}= & \left(\begin{array}{c}
0\\
0\\
0
\end{array}\right)\\
\boldsymbol{\hat{X}}_{\beta\alpha}= & \left(\begin{array}{c}
\sin\alpha\sin\beta\\
-\cos\alpha\sin\beta\\
0
\end{array}\right)\\
\boldsymbol{\hat{X}}_{\beta\alpha}= & \left(\begin{array}{c}
-\cos\alpha\cos\beta\\
-\sin\alpha\cos\beta\\
-\sin\beta
\end{array}\right)
\end{split}
\label{eq:d-tangent}
\end{equation}
It is clear that the torus is symmetric with the rotation about $z$ axis,  if we assume that
the nematic order has the same symmetry, then the $\theta$ will be function of $\beta$ only.
We consider this simple case first in the following. By substituting Eq.(\ref{eq:g^ij}), Eq.(\ref{eq:det-g}), Eq.(\ref{eq:d-nematic}) and Eq.(\ref{eq:d-tangent})
into the free energy functional (\ref{eq:3D-reform}) and considering that  $\theta_{\alpha}=0$ from symmetry assumption mentioned above, the free energy is derived as
\begin{eqnarray}
F[\theta]/K & = & \int\int d\alpha d\beta r\left(R+r\cos\beta\right)\nonumber \\
 &  & \cdot\left[\frac{\left(\frac{1}{4}\left(3-\text{cos}2\beta+2\text{cos}^{2}\beta\text{cos}2\theta\right)+2\text{sin}\beta\theta_{\alpha}+\theta_{\alpha}^{2}\right)}{\left(R+r\cos\beta\right)^{2}}+\frac{\text{sin}^{2}\theta+\theta_{\beta}^{2}}{r^{2}}\right]\label{eq:F-nonlinear}\\
 & = & \int\int d\alpha d\beta\frac{R+r\cos\beta}{r}\cdot\left[\theta_{\beta}^{2}+\sin^{2}\theta+\frac{r^{2}\cos^{2}\beta}{2\left(R+r\cos\beta\right)^{2}}\cos2\theta\right]+const\nonumber   \\
& = & \int\int d\alpha d\beta\frac{R+r\cos\beta}{r}\cdot\left[\theta_{\beta}^{2}+\left(1-\frac{r^{2}\cos^{2}\beta}{\left(R+r\cos\beta\right)^{2}}  \right)\sin^{2}\theta \right]+const\nonumber
\end{eqnarray}
%The free  energy is even in $\theta$, and can be expanded in powers of $\theta^2$ for small %$\theta$.  To the fourth power of $\theta$ the expansion is
%\begin{eqnarray}
%F[\theta]/K & = & \int\int d\alpha d\beta\frac{R+r\cos\beta}{r}\cdot\label{eq:ODE-Linear}\\
 %&  & \left[\theta_{\beta}^{2}+\left(1-\frac{r^{2}}%{\left(R+r\cos\beta\right)^{2}}\cos^{2}\beta\right)\theta^{2}+\frac{r^{2}}
%{3\left(R+r\cos\beta\right)^{2}}\theta^{4}\right]+const\nonumber
%\end{eqnarray}
%The coefficient of $\theta^{4}$ is always positive, if the coefficient of $\theta^2$ is also %positive, then $\theta=0$ is a minimum of the free energy.

Denote the ratio of the two radii as $k$,   $k\equiv r/R$, and the coefficient of $\sin^2\theta$ as $q(\beta)\equiv1-\frac{r^{2}}{\left(R+r\cos\beta\right)^{2}}\cos^{2}\beta$.  It is obvious
that the minimum of $q(\beta)$ is  $q(\pi)=1-\frac{1}{\left(k^{-1}-1\right)^{2}}$ at  $\beta=\pi$, and this minimum value is positive when $k<\frac12$. Thus,  $q(\beta)>0$ for all $\beta$ when $k<\frac12$ and
the free energy is a minimum when $\theta\equiv0$. As we will discuss shortly, this minimum is in fact the ground state of the system. When $k>\frac{1}{2}$,
$q\left(\pi\right)<0$,  then $\theta=0$ can not be the minimum free energy configuration. On the outer
circle,  where $\beta=0$ and $q(0) >0$, $\theta=0$ is the most favorable configuration,   i.e., the nematic  vector goes around the symmetry
axis. In the inner circle, for a very fat torus that the ratio $k$ closes to $1$,   The coefficient of $\sin^2\theta$ is negative in the region so that $\sin^2\theta$ takes its maximum value in order
to minimize the local free energy density, thus favors $\theta=\pi/2$, i.e. vector parallels to the symmetry axis.  From these observation it is clear that a constant $\theta$ could not minimize the total free energy. In order to find the solution of $\theta$ for these cases, we follow the standard procedure to find the result. Taking the
functional derivative of Eq.\ref{eq:F-nonlinear} and set it to zero, we obtain the following ordinary
differential equation
\begin{eqnarray}
\frac{\delta F[\theta]}{\delta\theta} & = & \frac{\partial f}{\partial\theta}-\frac{\partial}{\partial\beta}\left(\frac{\partial f}{\partial\theta'}\right)\nonumber \\
 & = & \frac{R+r\cos\beta}{r}\left(1-\frac{r^{2}\cos^{2}\beta}{\left(R+r\cos\beta\right)^{2}}\right)\sin2\theta\nonumber \\
 &  & +2\sin\beta\theta'-2\frac{R+r\cos\beta}{r}\theta''\nonumber \\
 & = & 0\label{eq:ODE}
\end{eqnarray}
where $f$ is the free energy density, and $\theta'$ and $\theta''$  denote the  first and second derivatives of $\theta$ with respect to
$\beta$, respectively.  There are two kinds of boundary conditions in this problem, the first one
is
\begin{equation}
\theta(0)=\theta(2\pi)\text{, }\theta'(0)=\theta'(2\pi)
\end{equation}
and the second one is
\begin{equation}
\theta(0)=0\text{, }\theta(2\pi)=\pi
\end{equation}
.

\begin{figure}[H]
\includegraphics{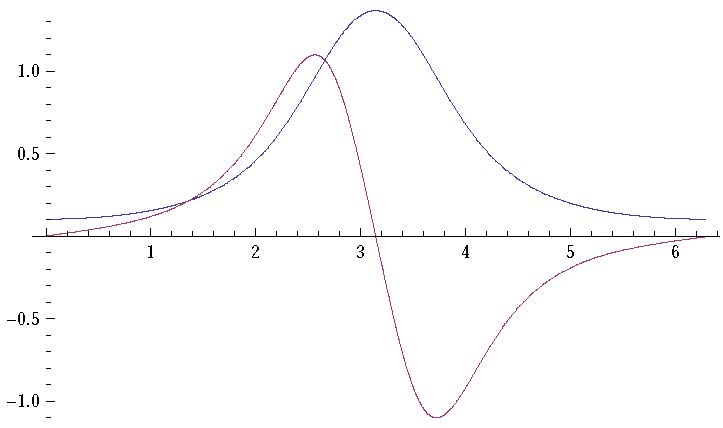}
\caption{variation of  $\theta$ (blue) and $\theta_\beta$(red)  with $\beta$   as calculated from the solution of (\ref{eq:ODE}) for $k=0.8$.  \label{fig:ODEk=00003D0.8}}
\end{figure}

With the first  boundary condition, the solution of $\theta$ is symmetric with $\pi$, i.e. $\theta(\beta)=\theta(2\pi -\beta)$, from this symmetry the first derivative of $\theta$ has to be
zero at $\beta=0$ and $\beta=\pi$.  With this condition, the boundary value problem (\ref{eq:ODE})  with boundary condition
\begin{equation}
\theta'(0)=\theta'(\pi)=0
\end{equation}
can be solved by the shooting method. Starting from $\beta=\pi$ for a guessed $\theta(\pi)$ and $\theta'(\pi)=0$, the equation can be integrated to $\beta=0$ to obtain $\theta(0)$ and $\theta'(0)$, adjusting the $\theta(\pi)$ according to the deviation of calculated $\theta'(0)$ from $0$ and integrate it again till convergence.  The calculated result for $k=0.8$ is shown
  In Fig.\ref{fig:ODEk=00003D0.8}.  In this case the calculated $\theta(\pi)=1.369226$,
$\theta(0)=0.107654$ respectively.

Figure \ref{fig:ODEk=00003D0.8} is the  variation with respect to $\beta$ the $\theta$ and
$\theta_\beta$ for $k=0.8$. On the outer circle where $\beta =0$ the $\theta$ is small so that
the nematic vector is only slightly away from the symmetry circle, it is increased monotonically
and reaches the maximum value at $\beta=\pi$, the inner circle.   The maximum value is slightly
smaller than $\pi/2$. This is typical for all $k$'s above a critical value $k_c$, below it a non-zero
solution to the equation (\ref{eq:ODE}) with BC1 does not exist.

\begin{figure}[tbh]
\includegraphics[width=0.5\textwidth]{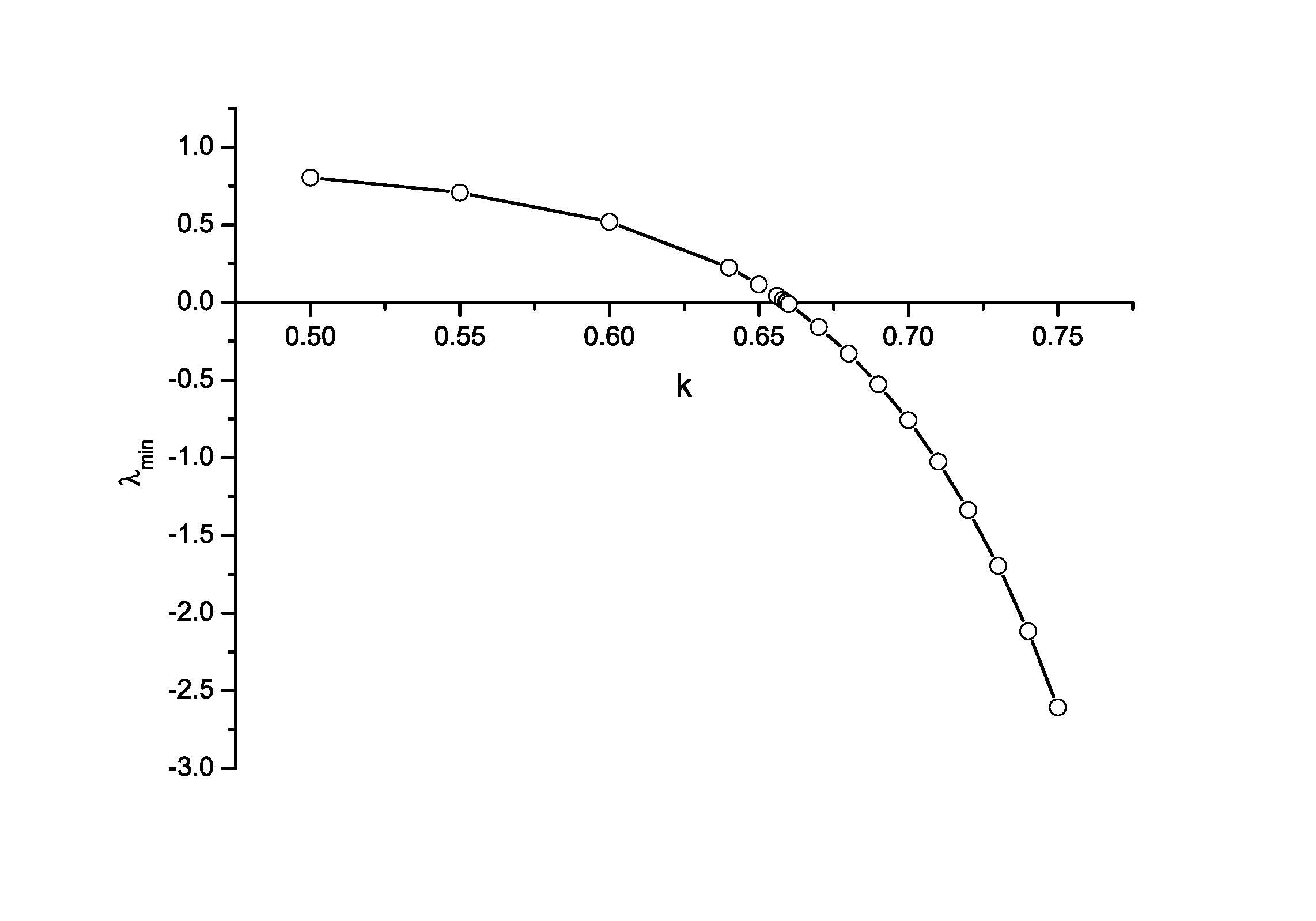}
\caption{The minimum eigenvalue $\lambda_\min$  as function of $k$. \label{fig:minimum-eigenvalue-}}
\end{figure}

In order to determine the  critical value $k_{c}$, we construct an eigenvalue problem with $\lambda$ as the eigenvalue to be determined:
\begin{equation}
-\theta''+\frac{\sin\beta}{\left(k^{-1}+\cos\beta\right)}\theta'+\frac{1}{2}\left(1-\frac{\cos^{2}\beta}{\left(k^{-1}+\cos\beta\right)^{2}}\right)\sin2\theta=\lambda\theta\label{eq:eigenODE}
\end{equation}
The spectrum of the $\lambda$ is a continuum with a lower bound, the minimum eigenvalue $\lambda_\min$.
If the minimum eigenvalue $\lambda_\min$ is positive, the only solution
 for the original equation is the constant one and when  $\lambda_\min$ is negative it is clear that there is a
non-constant solution.  The minimum eigenvalue is a function of $k$, and when $k=k_c$ the
minimum eigenvalue is zero. The minimum eigenvalue for different $k$ can be calculated
numerically by solving the eigenvalue equation (\ref{eq:eigenODE}), the result is shown in
 figure\ref{fig:minimum-eigenvalue-}. From the calculation the
critical value of $k$ is determined to be $k_{c}=0.659$.

\begin{figure}[H]
\includegraphics[width=0.5\textwidth]{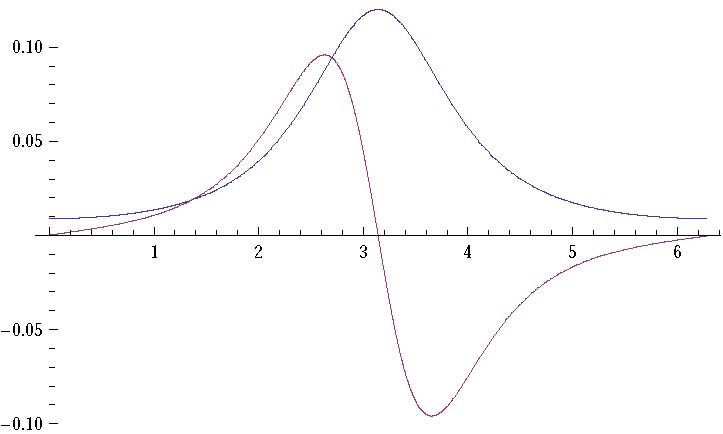}
\caption{The non-zero solution of $\theta$ for $k=0.66$, just above the $k_c=0.659$. The  blue line is the  $\theta(\beta)$ and  the red line is the $\theta'(\beta)$ respectively.\label{fig:ODEk=00003D0.66}}
\end{figure}

\begin{figure}[tbh]
\includegraphics[scale=0.4]{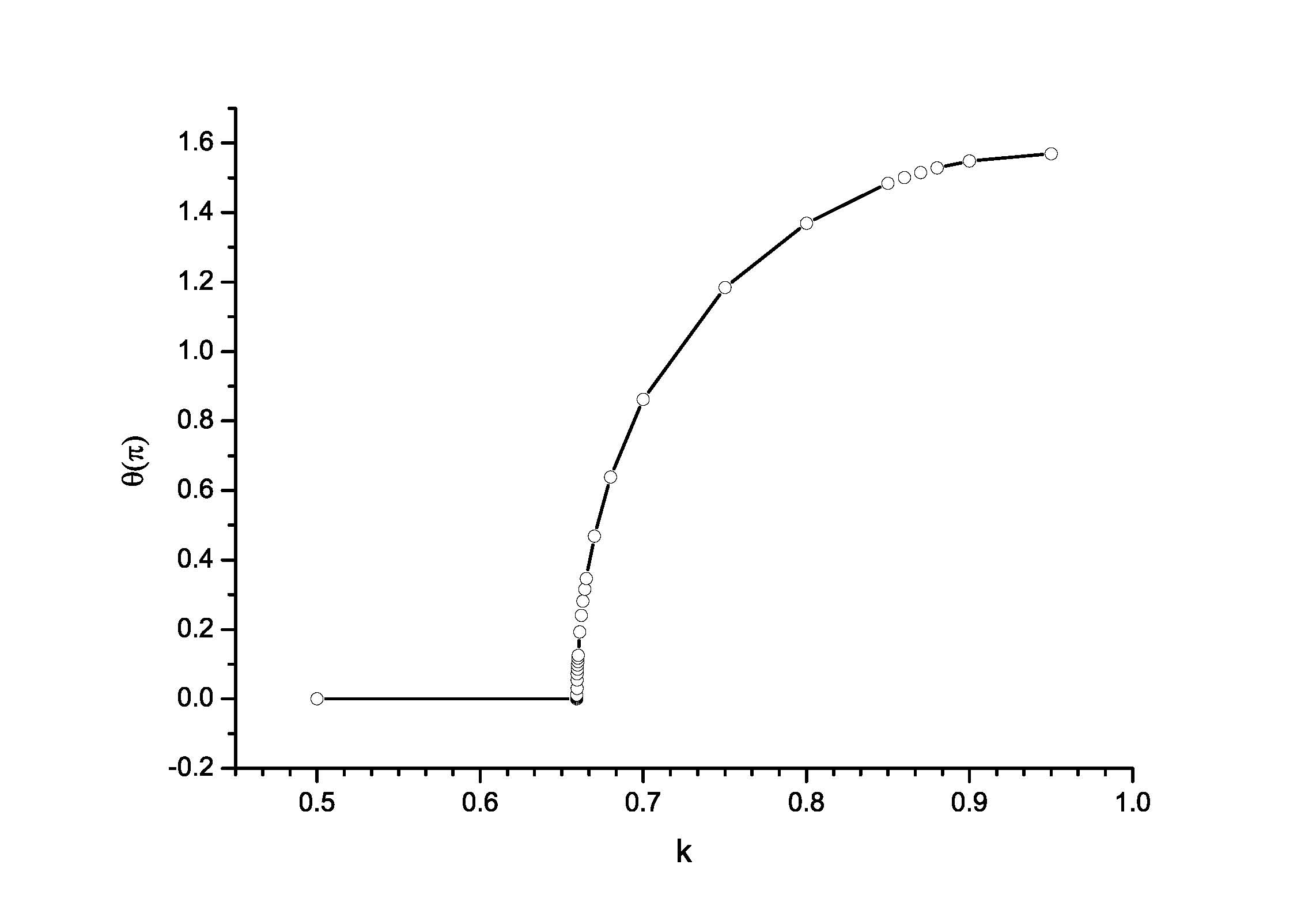}

\caption{Variation of the $\theta(\pi)$ with respect to $k$ in the vicinity of $k_{c}$\label{fig:theta(pi)}}
\end{figure}

The non-zero solution for the case that $k$ is  slightly above $k_{c}$,  $k=0.66$,
is given  in figure \ref{fig:ODEk=00003D0.66}. In this case,  $\theta(\pi)=0.1258247$ and  $\theta(0)=0.00965636$ respectively, which is very
small and close to zero.  Figure \ref{fig:theta(pi)} gave the value $\theta\left(\pi\right)$ as a function of $k$ in the vicinity of  the critical point $k_{c}$.  Below $k_{c}$,
$\theta\left(\pi\right) \equiv 0$  and $\theta\left(\pi\right)$  increases monotonically above $k_{c}$. By  a {log-log curve} fitting of the  data above $k_c$, we found that the $\theta(\pi)$
follows a power law form close to $k_c$ with the exponent $0.5$, i.e. $\theta(\pi)\sim\left(k-k_{c}\right)^{0.5}$.

\begin{figure}[tbh]
\includegraphics[width=0.5\textwidth]{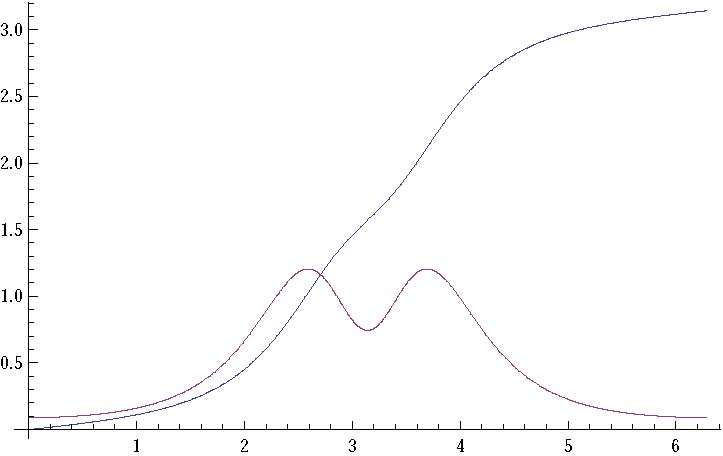}

\caption{k=0.8. Blue Line: $\theta(\beta).$ Red Line: $\theta'(\beta)$. with
$\theta(0)=0$,$\theta(2\pi)=\pi$\label{fig:k=00003D0.8BC2}}
\end{figure}

\begin{figure}[tbh]
\includegraphics[width=0.5\textwidth]{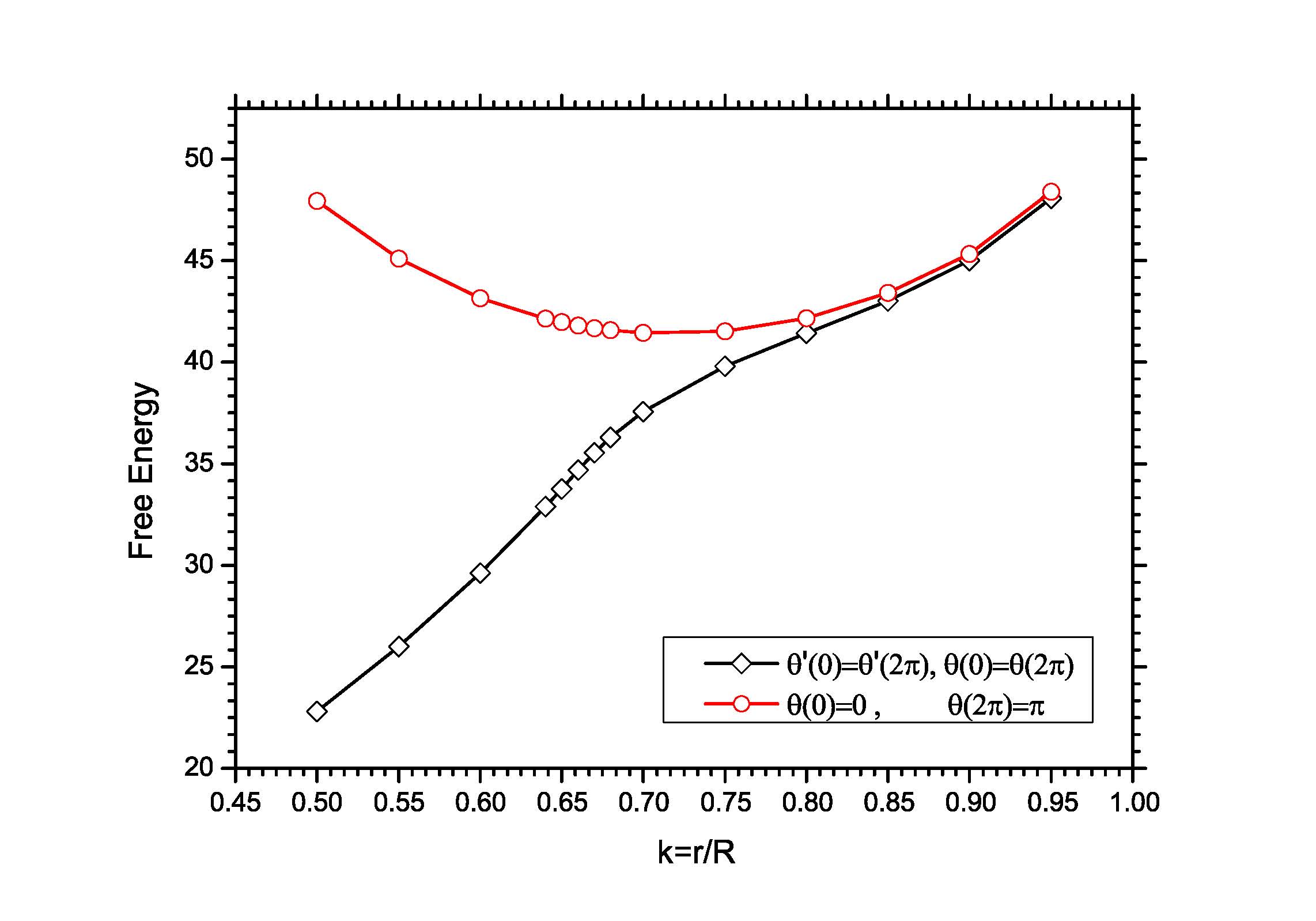}

\caption{Free energy for 2 sets of boundary conditions\label{fig:BC1BC2}}
\end{figure}

Next let us solve Eq.\ref{eq:ODE} with the second boundary condition (BC2)
\begin{equation}
\theta(0)=0\text{, }\theta(2\pi)=\pi
\end{equation}
At $k=0.8$, the solution is showed in Fig.\ref{fig:k=00003D0.8BC2}.
Note that its mirror solution is solution of Eq.\ref{eq:ODE} $\theta\left(0\right)=0$,
$\theta\left(2\pi\right)=-\pi$. It is clear the two chiral solutions
have the same free energy. We compare free energy of the solution
under BC1 and BC2 in Fig.\ref{fig:BC1BC2}. Free energy with BC1 always
lower than the one with BC2. When $k$ is large, the free energy difference
becomes very small. When $k$ is below $k_{c}$, zero-solution is the
ground state  and the free energy has an analytical form
\begin{equation}
F_{k}=\frac{4k\pi^{2}}{\sqrt{1-k^{2}}}
\end{equation}
For the solution with BC2, the free energy gets its minimum at $k\simeq0.72$.
Although the solution with BC2 has larger free energy than the one
with BC1, but it is topological stable if once formed unless thermal
fluctuate is large enough to destroy it wholly and cross the free energy barrier.

\begin{figure}[tbh]
\includegraphics[width=0.6\textwidth]{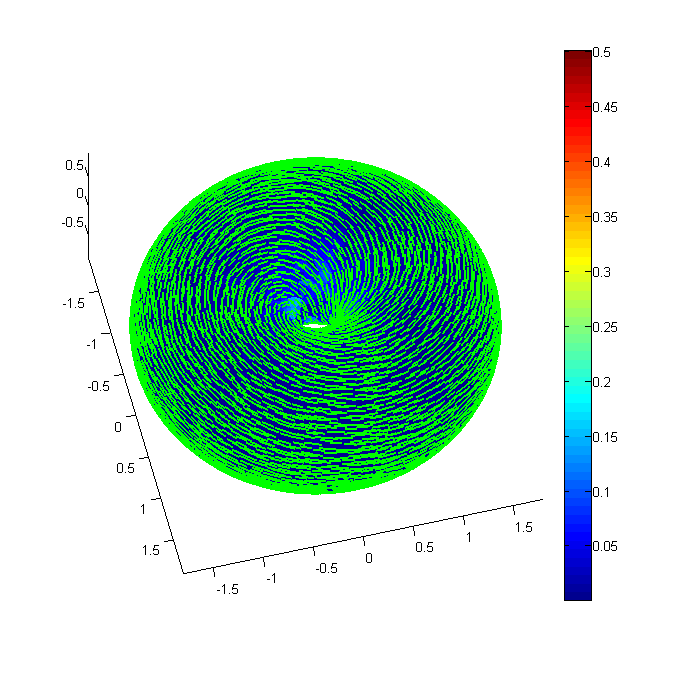}

\caption{Simulation result of defect-free state at $k=0.8$\label{fig:Simu-k0.8}}
\end{figure}

\begin{figure}[H]
\includegraphics[width=0.5\textwidth]{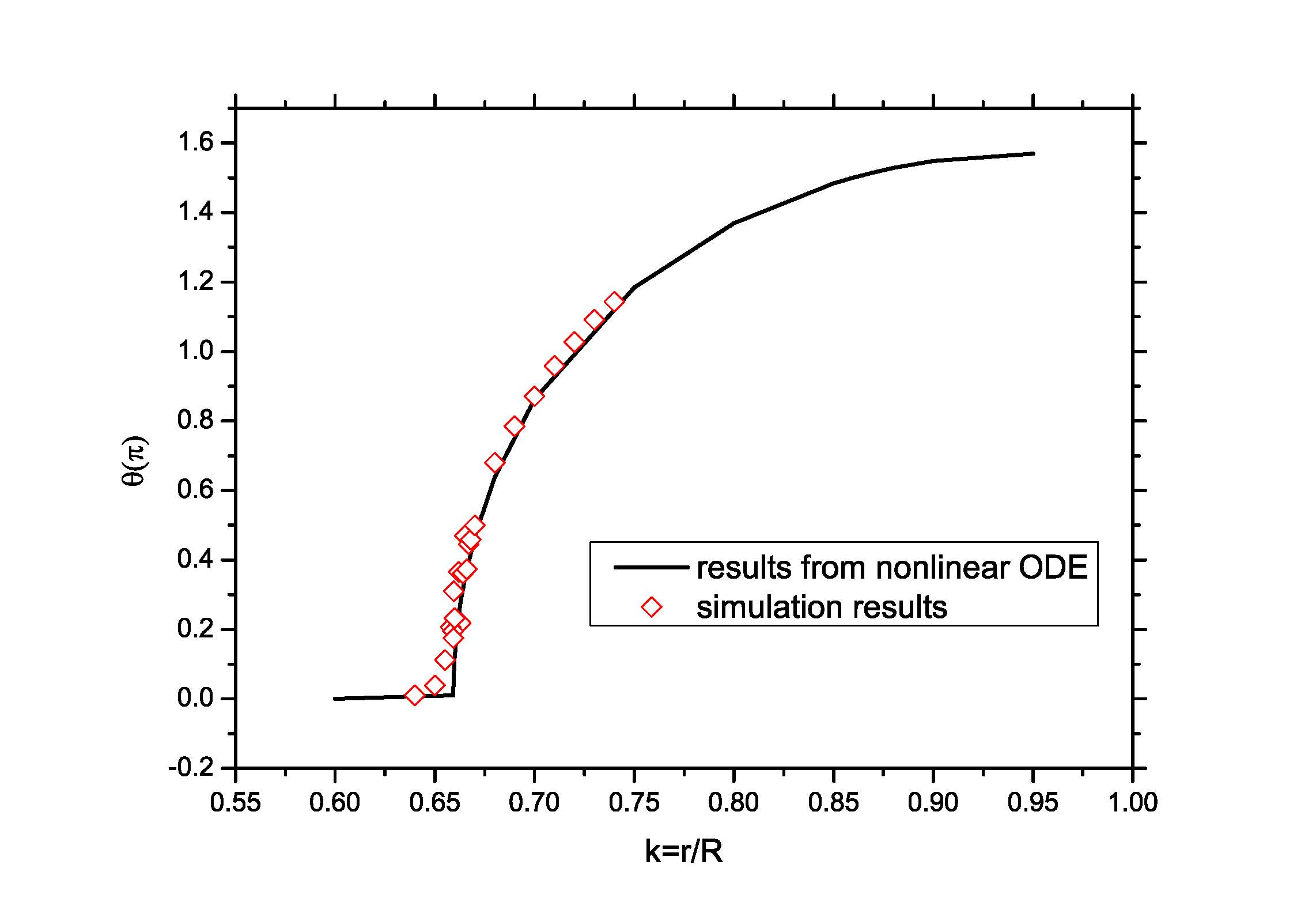}
\caption{$\theta(\pi)$ of the ground state, compare of simulation and analytical
results\label{fig:sim-anal}}
\end{figure}

Apart from the above analysis, we checked our results by the
 simulated annealing Monte Carlo calculation  for
the discrete free energy (\ref{eq:3D-discrete}). In order to give
a quantitative comparison, the total number of triangles of the surface mesh
is around $4\times10^{4}$. The initial temperature $k_{b}T/K$ is set to $1$,
then it is decreased by $T_{new}=T/1.0005$ every $10^{3}$
Monte Carlo steps. The simulated defect-free state at $k=0.8$
is  shown in figure \ref{fig:Simu-k0.8}, from which we confirm the axis-symmetry
condition is indeed true for this model. Figure \ref{fig:sim-anal} plots  $\theta\left(\pi\right)$
in the vicinity of $k_c$, the line is from the solution of the Eulerian equation and symbols are from simulation, the quantitative agreement  between the two are satisfactory.

\begin{figure}[tbh]
\includegraphics[width=0.5\textwidth]{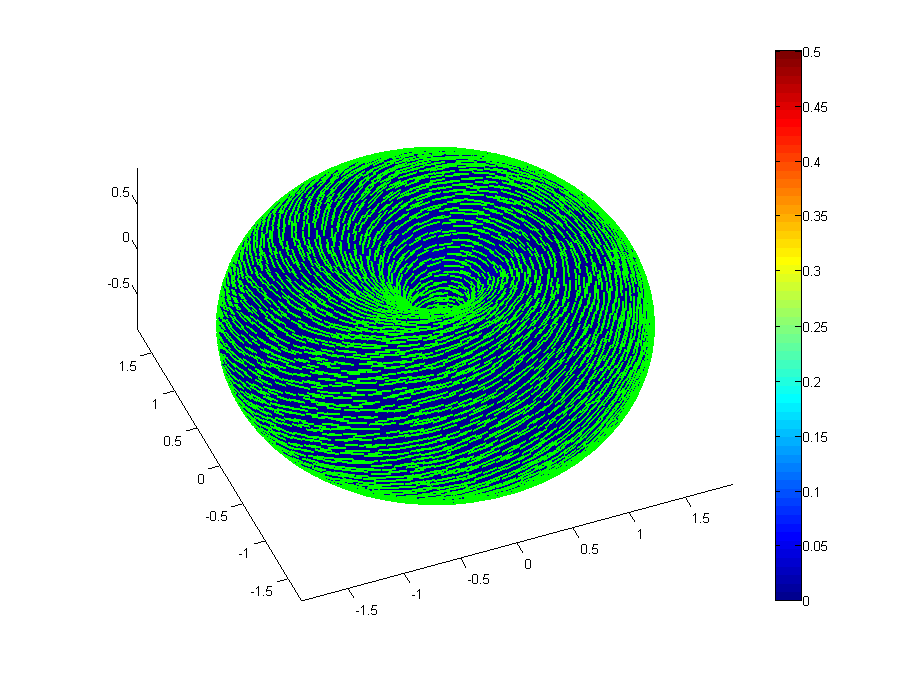}

\caption{Defect-free state of covariant gradient model at $k=0.8$\label{fig:covariant-k0.8}}
\end{figure}

Now we turn to the covariant gradient model (\ref{eq:Covarint-Continue}),
which can be rewritten as\citep{Nelson1987}

\begin{equation}
F_{Frank}=\int\left[K\left(\nabla\theta+\mathbf{A}\right)^{2}\right]dA\label{eq:covarint-explicit}
\end{equation}
where $\boldsymbol{A}=\left(\hat{X}_{\alpha}\cdot\partial_{\alpha}\hat{X}_{\beta}\text{, }\hat{X}_{\alpha}\cdot\partial_{\beta}\hat{X}_{\beta}\right)$
is the spin connect. The analysis and discussion of the  defect-free solutions of this model
was given in Ref.\citealp{Evans1995}. The defect-free states are
infinite continuum degeneracy configurations that $\theta$ equals
to an arbitrary constant angle everywhere. We checked here the results by using simulated annealing Monte Carlo simulation with discrete model Eq.\ref{eq:covariant-discrete}.
Our calculation indicates  that the orientation angle $\theta$ is indeed a constant with
random value which confirms the result in\citealp{Evans1995}.    A representative state at $k=0.8$ are
shown in Fig.\ref{fig:covariant-k0.8}.  It is clear that the defect-free state of covariant
gradient model is very different from 3D gradient model.  To clarify this difference, we rewritten the 3D gradient
model Eq.\ref{eq:Covarint-Continue}  as \citep{Selinger2011}
\begin{equation}
F_{Frank}=\int\left[K\left(\nabla\theta+\mathbf{A}\right)^{2}+K\left(\hat{n}\cdot\mathbf{K}\cdot\mathbf{K}\cdot\hat{n}\right)\right]dA\label{eq:3D-reform}
\end{equation}
where $\mathbf{K}$ is the curvature tensor. Compared to Eq.\ref{eq:covarint-explicit},
Eq.\ref{eq:3D-reform} has the additional term $\hat{n}\cdot\mathbf{K}\cdot\mathbf{K}\cdot\hat{n}$,
which is the extrinsic coupling between nematic order and the curvature
tensor, favoring aligning the nematic order along small curvature
direction.

\subsection{Topological defects}

Now we turn to the more interesting case of   topological defects. Defect-free states is the
simplest case that the Poincar\'{e}-Hopf theorem can be satisfied, however states with topological
order can also satisfy the Poincar\'{e}-Hopf theorem provided that the defects appear in pairs
so that the total $\chi=0$.

Headless nematic vector is a 2-atic order, which has the $\pi$ rotational
symmetry. Thus the winding numbers of least energetically expensive
topological defects in nematic field are $\pm\frac{1}{2}$.  When defects
presents, the symmetry of rotation around $z$ axis is no longer exist,
thus the analysis through the Euler equation is much harder than the
defect-free case. We do not attempt to perform such an analysis, instead,
we use the simulated annealing Monte Carlo method to investigate. The
computation load in the defect-free case and the case with topological
defects is basically the same and the algorithm is reliable based on the
simulation of the defect-free problem which gave the quantitative correct
results.

\begin{figure}[tbh]
\includegraphics[width=0.45\textwidth]{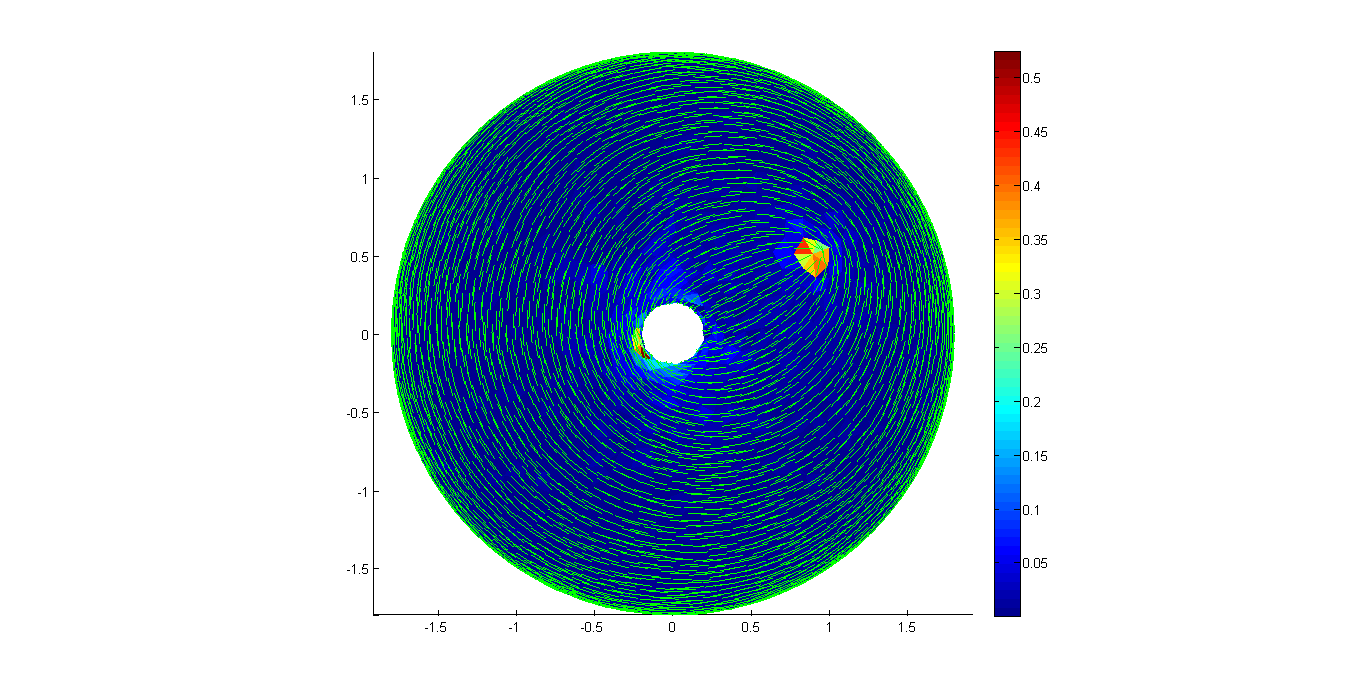}\includegraphics[width=0.45\textwidth]{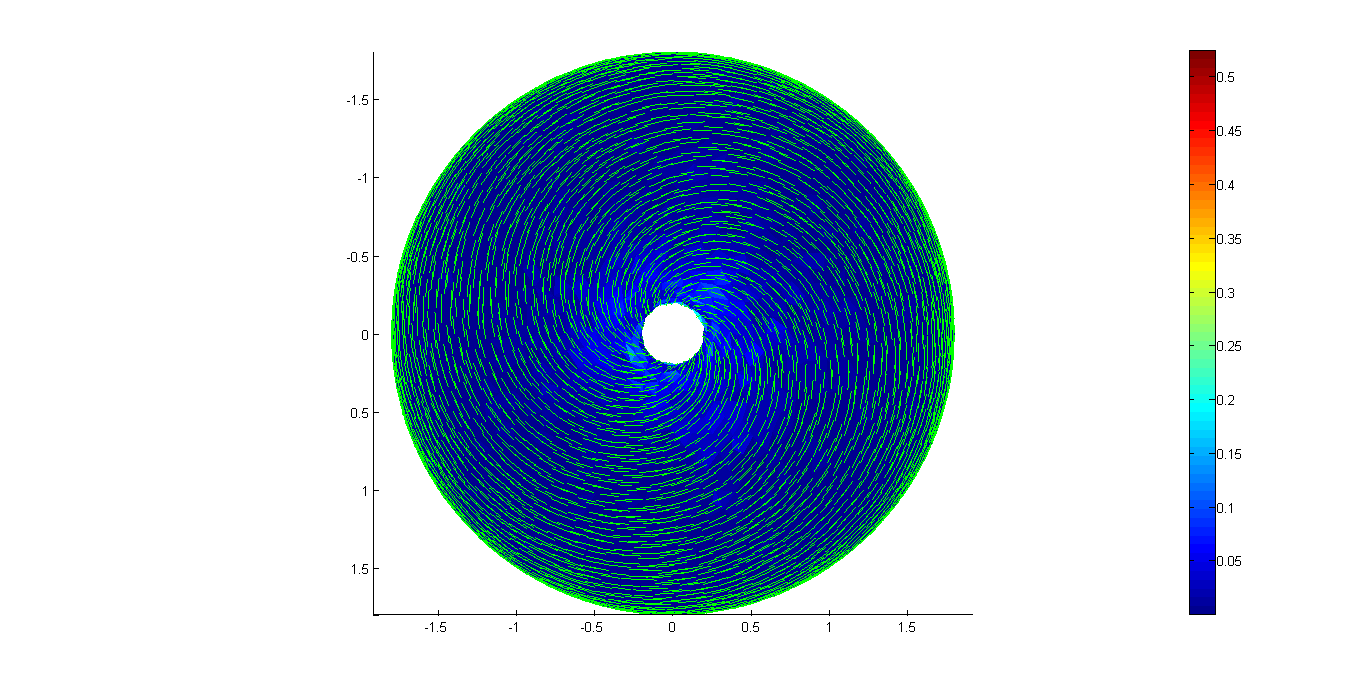}

\caption{The configuration with a pair of defects for direct derivative model, top view(left) and bottom view (right) \label{fig:Direct-3D}}
\end{figure}

\begin{figure}[tbh]
\includegraphics[width=0.5\textwidth]{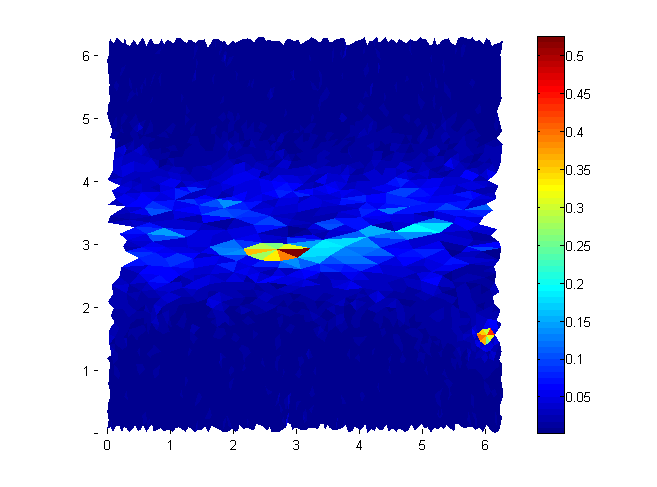}

\caption{Projection to the $\alpha -\beta$ plane of the configuration in figure \ref{fig:Direct-3D}, the two defects are clearly seen. \label{fig:Direct-2D}}
\end{figure}

\begin{figure}[tbh]
\includegraphics[width=0.5\textwidth]{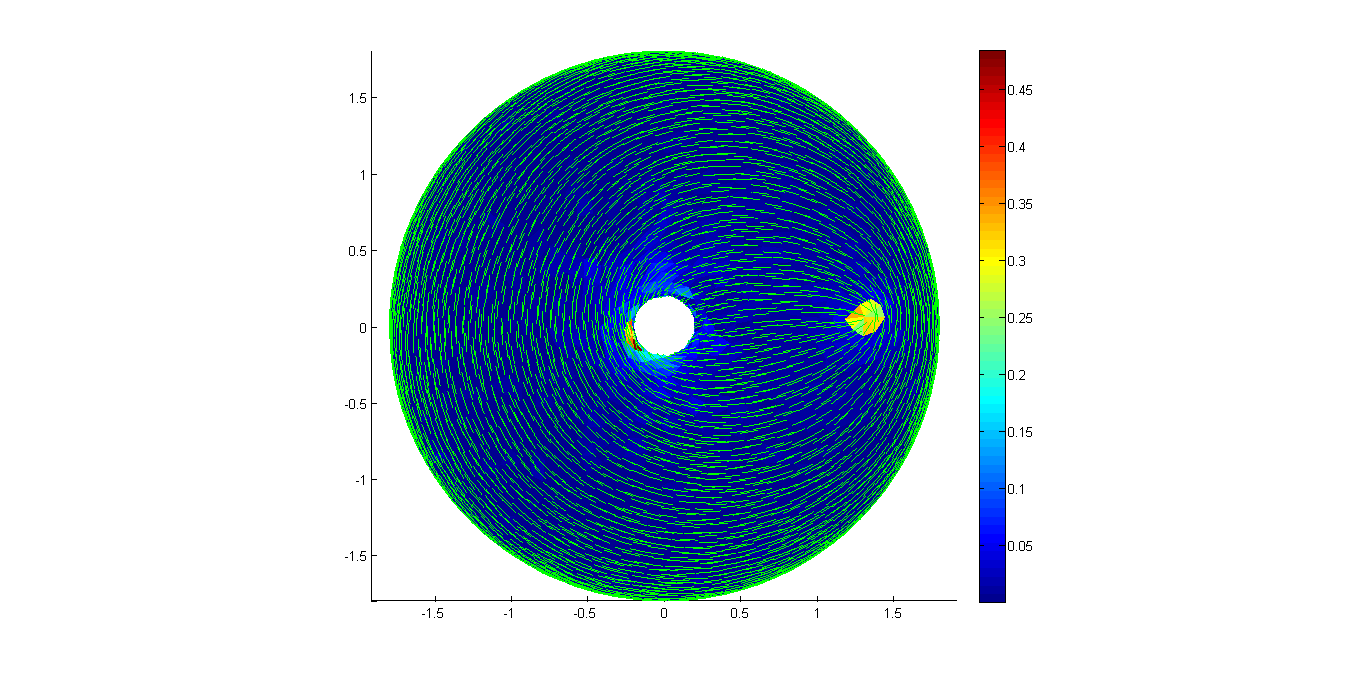}\includegraphics[width=0.59\textwidth]{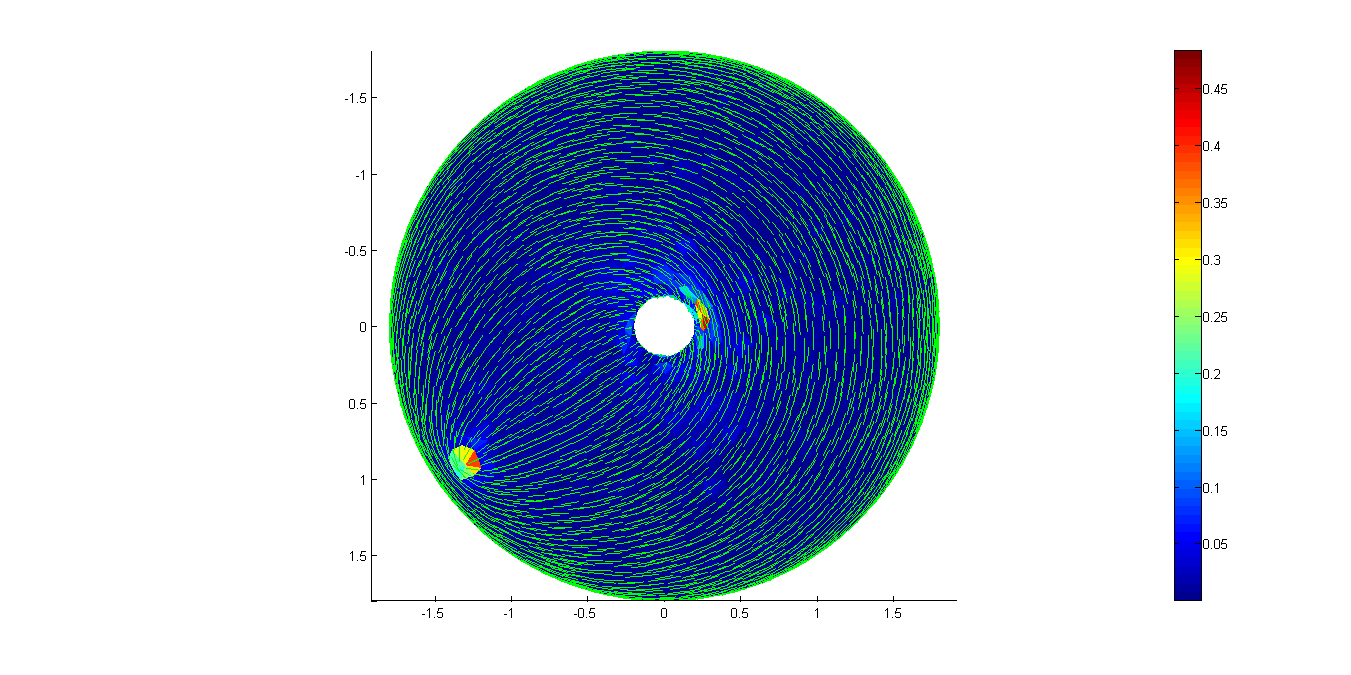}

\caption{The configuration with two pairs of defects for direct derivative model    \label{fig:Direct-2pair-3D}}

\end{figure}

\begin{figure}[tbh]
\includegraphics[width=0.5\textwidth]{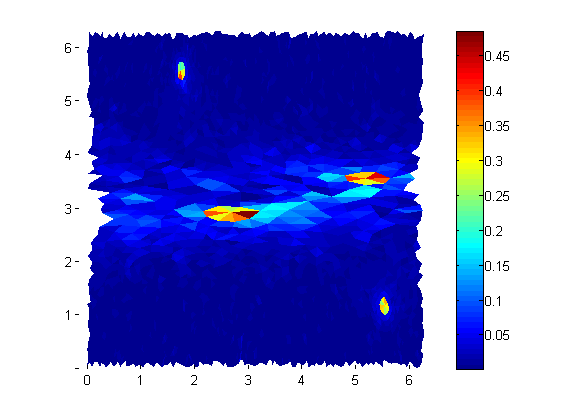}

\caption{Projection to the $\alpha-\beta$ plane of the configuration in figure \ref{fig:Direct-2pair-3D}, the four defects are clearly seen in the figure.\label{fig:Direct-2pair-2D}}

\end{figure}

Figure  \ref{fig:Direct-3D} shows a typical configuration with one pair
of topological defects for $k=0.8$.  and Figure \ref{fig:Direct-2D}
shows its 2D projection to the $\alpha-\beta$ plane.  As we can see that the  $+1/2$
defect is  located on the outer of the torus where the Gaussian curvature
is positive, and the $-1/2$ defect  is  located at the inner of the
torus where Gaussian curvature is the most negative.
It is known that  the topological defects on curved membrane behaviors in a way
similar to an electrostatic problem. In this analogy  the  positive
Gaussian curvature  plays the rule of a negative charge distribution,
and positive topological defect plays the rule of  a positive point
charge{[}Nelson Review{]}. So the high positive Gaussian curvature
geometry attracts positive topological defects, and vise versa.  The pair of
defects is in fact a direct result from simulation, i.e., when we  starting the calculation
from a random configuration and lower the temperature subsequently, both  defect-free
state or the state with defects may appear.  The configuration of two
pairs of defects   are also observed, which is shown
in Fig.\ref{fig:Direct-2pair-3D}, and its $2D$ projection to $\alpha-\beta$
plane is shown in Fig. \ref{fig:Direct-2pair-2D}. The two $+1/2$ defects are located
on the outer, and the two $-1/2$ defects located on the inner, as expected from
the electrostatic analogy.

The ground state is the defect-free state and the state with topological defects are
the excited states. When the positive and negative defects of a defect  pair is close enough
the two will annihilate, however, there is a potential barrier between the two defects which
prevent  the two close to each other. The barrier height determines the life time of the excited
states. For fat torus the barrier height is high  so that the life time of the states with
defects is long,  while  for  thin torus of smaller $k$, the barrier is low  and the life time
of the excited state is short. In the simulation calculations defects are  not observed for thin
torus.

\begin{figure}[tbh]
\begin{minipage}{0.49\textwidth}
\includegraphics[width=0.9\columnwidth]{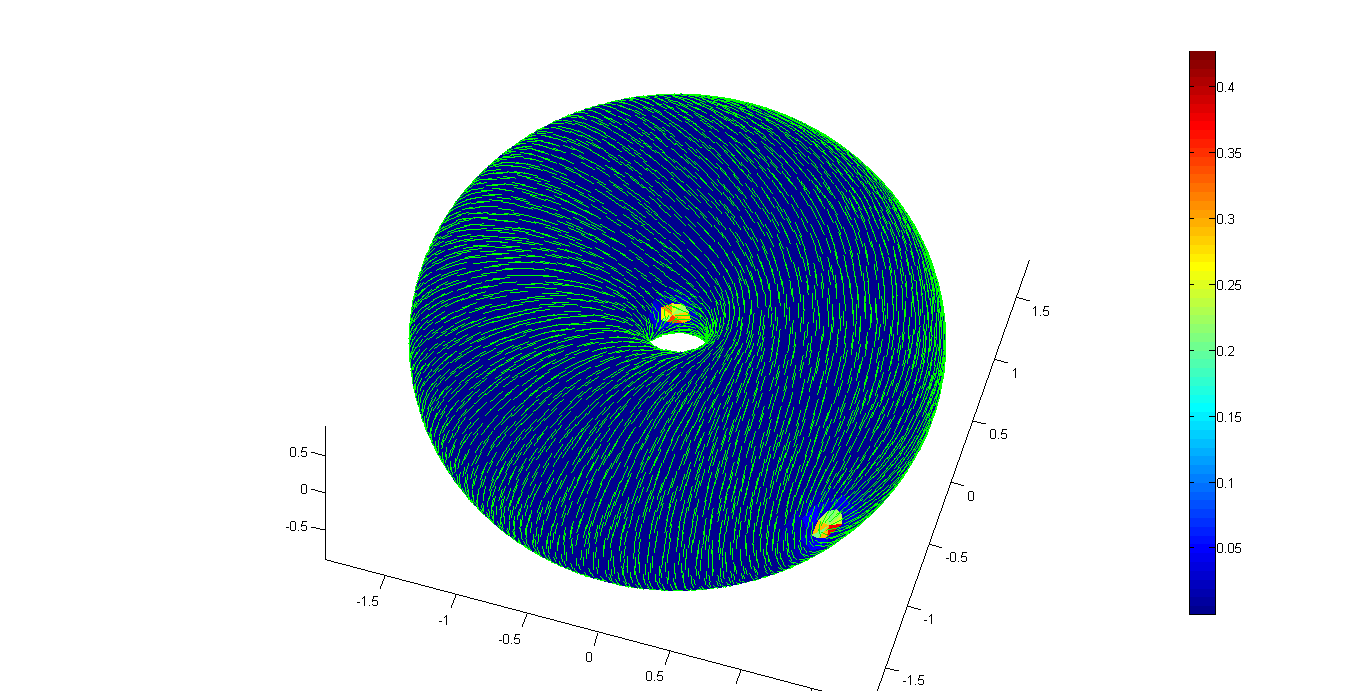}

\caption{2 pair of defects, k=0.8, direct dot\label{fig:2-pair-3d}}
%\end{figure}
\end{minipage} \begin{minipage}{0.49\textwidth}

%\begin{figure}[tbh]
\includegraphics[width=0.9\columnwidth]{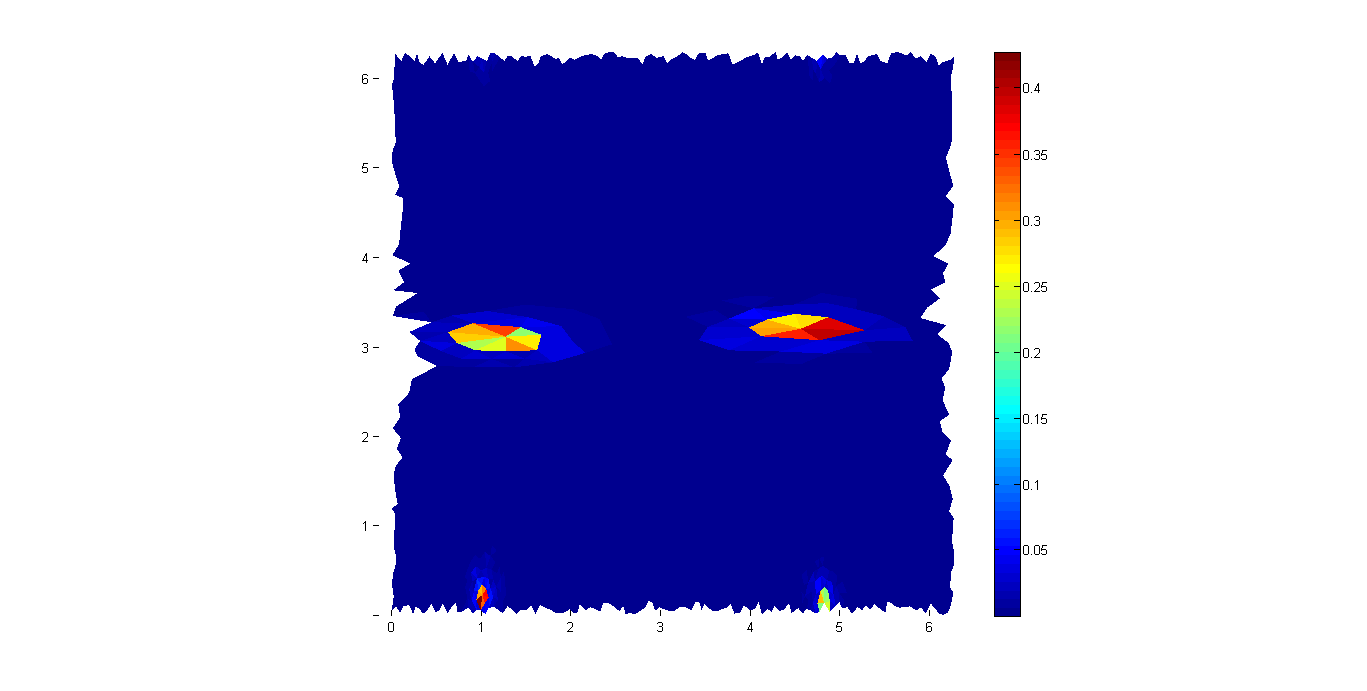}

\caption{2 pair of defects, k=0.8, $2D$ projection\label{fig:2-pair-3d-1}}
\end{minipage}
\end{figure}

\begin{figure}[tbh]
\includegraphics[width=0.5\textwidth]{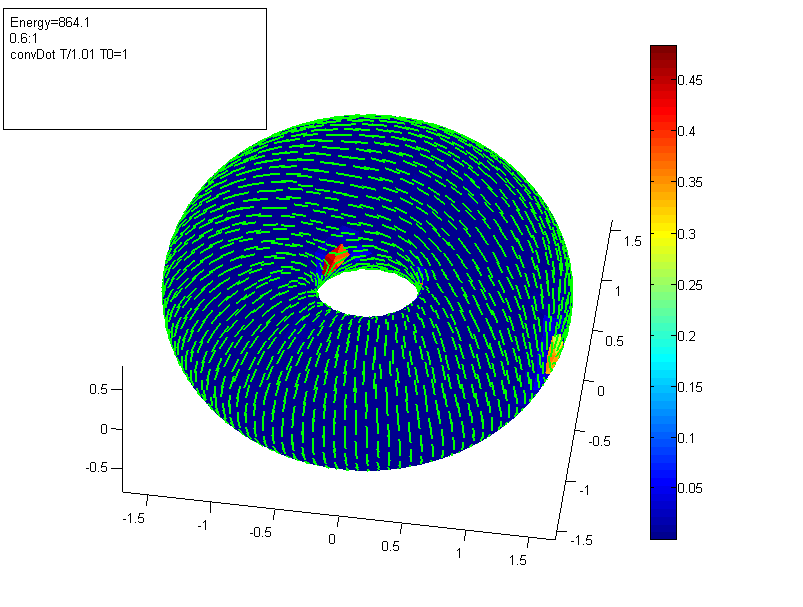}

\caption{2 pair of defects, k=0.6\label{fig:cov-0.6}}

\end{figure}

For the covariant derivative model, a  typical
configuration with two pairs of defects is shown in figure \ref{fig:2-pair-3d},  along with
its $2D$
projection to $\alpha-\beta$ plane is shown in figure \ref{fig:2-pair-3d-1}.
Similar to the $3D$ direct derivative model,  the two $+1/2$ defects
are located on the outer part of the torus, and the two $-1/2$ defects are located on the
inner part.  A striking feature of this model compared to the previous one is that
all the defects are located at the middle plane of the torus ($\beta=0,\pi$). This  is
a result of the  mirror
symmetry about the middle plane of the orientation field.  The second difference
is that for a relative thin torus at $k=0.6$,
we also observed the configuration of two pairs of topological defects as
shown in  figure \ref{fig:cov-0.6}, while for a $3D$ direct derivative
model no configuration with defects  is observed at $k=0.6$. This results
are consistent with the theoretically predicted ground eigenstates
$2_{1}$ by Evans\cite{Evans1995}.

\section{Conclusion\label{sec:Conclusion}}
In summary, we calculated the nematic ordering on a torus with both the standard and 
covariant Frank free energy models. We found that the ground state of the standard model
is $\theta=0$ for small $k$ and $\theta$ various with $\beta$ for larger $k$. The transition
point $k_c=0.659$ is determined numerically from an eigenvalue problem.  It is found that $\theta(\pi)$ changes continuously from $0$ to non-zero at $k_c$ in a power law form and 
the exponent was determined to be $0.5$, i.e., $\theta(\pi) \sim (k-k_c)^{0.5}$ for $k$ close
to $k_c$. We also calculated the nematic configuration for $k$ above $k_c$. All these results
were obtained from the solution of the Eulerian equation of the free energy functional and
checked by the Monte Carlo simulated annealing method. In the case of the covariant model, 
we confirmed the ground state is the $\theta$ equals to a random constant, which was proposed
by Evans in  Ref.\citealp{Evans1995}. The excited states are the states with pairs of defects.
In the simulation, we observed one pair and two pairs defective states in both models.
The positive defects are located on the outer part of the torus and the negative ones are
on the inner part.The results can be understood by the analogy with an electrostatic problem where the Gauss curvature plays the rule of a continuous negative charge distribution. The excited state, when created, can stay for very long time because the barrier for the annihilation of a defect pair is fairly high for fat tori. This means that the defect state
is experimentally observable and we hope that future experiment will reveal such state.

Work supported by the National Nature Science Foundation of China under grant \#10874111,  \#11304169 and
\#11174196.

\bibliographystyle{rsc}
\bibliography{LCtorus}

\end{document}